\documentstyle[12pt]{article}
\begin{document}
\centerline{\Large \bf Negative Energy Density States for the Dirac Field}
\centerline{\Large \bf in Flat Spacetime}
\vskip .9in
\centerline{Dan N. Vollick}
\centerline{Department of Physics and Astronomy}
\centerline{University of Victoria}
\centerline{Victoria, British Columbia}
\centerline{P.O. BOX 3055 MS7700}
\centerline{Canada}
\centerline{V8W 3P6}
\vskip .9in
\section*{\centerline{\bf Abstract}}
Negative energy densities in the Dirac field produced by state vectors
that are the superposition of two single particle electron states are
examined. I show that for such states the energy 
density of the field is not bounded from
below and that the quantum inequalities derived for scalar fields are
satisfied. I also show that it is not possible to produce negative
energy densities in a scalar field using state vectors that are 
arbitrary superpositions of single particle states.
\section*{Introduction}
Recent work on wormholes \cite{Mo1,Mo2,Fr1} and the `warp drive'
\cite{Al1} has generated interest in matter
that violates the weak energy condition. Most discussions of such exotic 
matter occur within the context of quantum field theory and deal with bosonic
fields
\cite{Mo1,Mo2,Ep1,Ku1,Fo1,Fo2} (see \cite{Vo1,Vo2,Vo3} for a classical discussion
and \cite{Gr1} for a discussion of fermionic fields in a curved spacetime).
Recently Ford and Roman \cite{Fo1,Fo2} have shown that 
in flat spacetime the energy density
of a massless scalar field and the electromagnetic field satisfy the
quantum inequalities
\begin{equation}
\hat{\rho}\equiv\frac{t_0}{\pi}\int_{-\infty}^{\infty}\frac{<:T_{00}:>}{
(t^2+t_0^2)}dt\geq -\frac{A}{2V}\sum_k\omega_ke^{-2\omega_k t_0}
\label{QI1}
\end{equation}
in the finite volume case and
\begin{equation}
\hat{\rho}\geq -\frac{3A}{32\pi^2t_0^4}
\label{QI2}
\end{equation}
in the infinite volume case, 
where $<:T_{00}:>$ is the expectation value of the normal ordered
energy density, $A=1$ for a massless scalar field, and $A=2$
for the electromagnetic field. The quantity $\hat{\rho}$
samples $<:T_{00}:>$ over a time of order $t_0$. For simplicity I will
refer to $<:T_{00}:>$ as the energy density of the field.

Unfortunately the methods used above to obtain general constraints on 
the energy densities cannot be applied to the Dirac equation. 
In this paper
I will look at the negative energy densities that occur in
states that are the superposition of two single particle electron states.
For such states I show that the energy density is not bounded from
below and that 
an observer at a fixed spatial point sees the energy density as a wave
propagating by at the speed of light superimposed on a positive
background.
In certain regions of this wave the energy density
is negative and in other regions it is positive. Thus
the negative energy densities do not persist indefinitely. In fact,
I will show that the negative energy density persists for a time that is 
inversely proportional to the minimum value of the energy density.
I will also show that the quantum inequality (\ref{QI1}) is satisfied
for the states considered in this paper.
   
Finally I will show that, in contrast to the Dirac field, a scalar field
cannot have negative energy densities for states that are arbitrary
superpositions of single particle states.
   
Throughout this paper I will take $\hbar=c=1$ and the metric will be
taken to have the signature (-+++).
\section*{Negative Energy States for the Dirac Field}
The Lagrangian for the Dirac field is
\begin{equation}
L=\frac{\imath}{2}\bar{\psi}\gamma^{\mu}\stackrel{\leftrightarrow}
{\partial}_{\mu}\psi
-m\bar{\psi}\psi .
\end{equation}
Since the canonical energy momentum tensor $\theta^{\mu\nu}$ is not symmetric
the Belinfante tensor \cite{Be1}
\begin{equation}
\begin{array}{ll}
T^{\mu\nu}=\theta^{\mu\nu}-\frac{\imath}{2}\partial_{\alpha}[\frac
{\partial L}{\partial(\partial_{\alpha}\psi^l)}(J^{\mu\nu})^l_{\;\;m}
\psi^m+\frac{\partial L}{\partial (\partial_{\alpha}\bar{\psi}^l)}
(\bar{J}^{\mu\nu})^l_{\;\;m}\bar{\psi}^m-\frac{\partial L}{\partial
(\partial_{\mu}\psi^l)}(J^{\alpha\nu})^l_{\;\; m}\psi^m\\
-\frac{\partial L}{\partial(\partial_{\mu}\bar{\psi}^l)}(\bar{J}^{\alpha
\nu})^m_{\;\; m}\bar{\psi}^m-\frac{\partial L}{\partial(\partial_{\nu}
\psi^l)}(J^{\alpha\mu})^l_{\;\; m}\psi^m-\frac{\partial L}{\partial(
\partial_{\nu}\bar{\psi}^l)}(\bar{J}^{\alpha\mu})^l_{\;\; m}\bar{\psi}^m]
\end{array}
\end{equation}
should be used, where 
$J^{\mu\nu}$ is the generator of Lorentz 
transformations for $\psi$ and, $\bar{J}^{\mu\nu}$ is the generator of
Lorentz transformations for $\bar{\psi}$. A short calculation gives
\begin{equation}
T_{\mu\nu}=\frac{\imath}{4}[\bar{\psi}\gamma^{\mu}
\stackrel{\leftrightarrow}{\partial}^{\nu}\psi
+\bar{\psi}\gamma^{\nu}\stackrel{\leftrightarrow}
{\partial}^{\mu}\psi].
\end{equation}
Thus
\begin{equation}
T_{00}=\frac{\imath}{2}[\psi^{\dagger}\dot{\psi}-\dot{\psi}^{\dagger}\psi].
\end{equation}
Now define
\begin{equation}
<\rho>= <:T_{00}:>
\label{rho}
\end{equation}
and consider the Dirac field in a box of volume $V$.
The field operator can be written in terms of creation and annihilation
operators as
\begin{equation}
\psi(x)=\sum_k\sum_{\alpha=1,2}[b_{\alpha}(k)u^{\alpha}(k)e^{
\imath k\cdot x}+d^{\dagger}_{\alpha}(k)v^{\alpha}(k)e^{-\imath k\cdot x}]
\label{psi}
\end{equation}
where 
\begin{equation}
u^{\alpha}(k)=\left (
\begin{array}{ll}
\sqrt{\frac{\omega+m}{2\omega V}}\phi^{\alpha}\\
\frac{\vec{\sigma}\cdot \vec{k}}{\sqrt{2\omega(\omega +m)V}}\phi^{\alpha}
\end{array}
\right ),
\label{u}
\end{equation}
\begin{equation}
v^{\alpha}(k)=\left (
\begin{array}{ll}
\frac{\vec{\sigma}\cdot \vec{k}}{\sqrt{2\omega(\omega +m)V}}\phi^{\alpha}\\
\sqrt{\frac{\omega +m}{2\omega V}}\phi^{\alpha}
\end{array}
\right ),
\end{equation}
$\phi^{1\dagger}=(1,0)$, and $\phi^{2\dagger}=(0,1)$.
The creation and annihilation operators satisfy
\begin{equation}
\{b_{\alpha}(k),b^{\dagger}_{\alpha^{'}}(k^{'})\}=\delta_{\alpha,\alpha^{'}}
\delta_{k,k^{'}}
\end{equation}
and
\begin{equation}
\{d_{\alpha}(k),d^{\dagger}_{\alpha^{'}}(k^{'})\}=\delta_{\alpha,\alpha^{'}}
\delta_{k,k^{'}}
\end{equation}
with all other anticommutators vanishing. Substituting (\ref{psi}) 
into (\ref{rho}) gives
\begin{equation}
\begin{array}{ll}
<\rho>=\frac{1}{2}\sum_{k,k^{'}}\sum_{\alpha,\alpha^{'}}\{
(\omega_k+\omega_{k^{'}})[<b^{\dagger}_{\alpha}(k)b_{\alpha^{'}}(k^{'})>u^{
\dagger\alpha}(k)u^{\alpha^{'}}(k^{'})e^{-\imath (k-k^{'})\cdot x}+\\
<d^{\dagger}_{\alpha^{'}}(k^{'})d_{\alpha}(k)>v^{\dagger\alpha}(k)v^{\alpha^{'}}
(k^{'})e^{\imath (k-k^{'})\cdot x}]\}
+(\omega_{k^{'}}-\omega_k)[<d_{\alpha}(k)b_{\alpha^{'}}(k^{'})>
v^{\dagger\alpha}(k)u^{\alpha^{'}}(k^{'})e^{\imath (k+k^{'})\cdot x}\\
-<b^{\dagger}_{\alpha}(k)d^{\dagger}_{\alpha^{'}}(k^{'})>u^{\dagger\alpha}(k)v^{\alpha
^{'}}(k^{'})e^{-\imath (k+k^{'})\cdot x}]\}
\end{array}
\label{rho2}
\end{equation}
where I have used $:d_{\alpha}(k)d_{\alpha^{'}}(k^{'}):=-d_{\alpha
^{'}}^{\dagger}(k^{'})d_{\alpha}(k)$.
   
Now consider a state vector of the form
\begin{equation}
|\psi>=\frac{1}{\sqrt{1+\lambda^2}}[|k_z,1>+\lambda |k_x,2>]
\label{state}
\end{equation}
where $|k,\alpha>=b^{\dagger}_{\alpha}(k)|0>$ and $\lambda$ is real. Since the
state $|\psi>$ contains only electrons all expectation values in 
(\ref{rho2}) containing $d_{\alpha}(k)$ or $d^{\dagger}_{\alpha}(k)$ vanish.
Substituting (\ref{state}) and (\ref{u}) into (\ref{rho2}) gives
\begin{equation}
<\rho>=\frac{1}{(1+\lambda^2)V}[\omega_{k_z}+\lambda
\beta+\lambda^2\omega_{k_x}]
\end{equation}
where
\begin{equation}
\beta=\frac{k_xk_z(\omega_{k_x}+\omega_{k_z})\cos\theta}{2\sqrt{
\omega_{k_x}\omega_{k_z}(\omega_{k_x}+m)(\omega_{k_z}+m)}}
\end{equation}
and $\theta=(k_x-k_z)x$. Note that $<\rho>=\omega_{k_z}/V$
for $\lambda=0$ and $<\rho>=\omega_{k_x}/V$ as $\lambda\rightarrow
\infty$, as expected. It is easy to see that $<\rho>$ will be 
negative if
\begin{equation}
\beta^2>4\omega_{k_x}\omega_{k_z}
\label{beta}
\end{equation}
and if
\begin{equation}
-\frac{\beta}{2}-\sqrt{\left(\frac{\beta}{2}\right)^2-\omega_{k_x}\omega_{k_z}}
<\omega_{k_x}\lambda<-\frac{\beta}{2}+\sqrt{\left(\frac{\beta}{2}\right)^2-
\omega_{k_x}\omega_{k_z}} .
\label{lambda}
\end{equation}
Consider the ultrarelativistic limit, 
$k_x,k_z>>m$. In this limit
\begin{equation}
\beta =\frac{1}{2}(\omega_{k_x}+\omega_{k_z})\cos\theta.
\end{equation}
Equation (\ref{beta}) becomes
\begin{equation}
\cos^2\theta>\frac{16\omega_{k_x}\omega_{k_z}}{(\omega_{k_x}+
\omega_{k_z})^2} .
\end{equation}
For a solution to exist it is necessary that $16\omega_{k_x}
\omega_{k_z}\leq (\omega_{k_x}+\omega_{k_z})^2$. This will be 
satisfied if $\omega_{k_x}\geq (7+\sqrt{48})\omega_{k_z}$ or if
$\omega_{k_z}\geq (7+\sqrt{48})\omega_{k_x}$. Thus it is possible
to produce negative energy densities for state vectors of the form
(\ref{state}) if $\lambda$ is chosen to satisfy (\ref{lambda}).
  
It is now easy to show that the energy density is not bounded from
below. To simplify the discussion take $x^{\mu}=0$, so that $\cos
\theta=1$. In the ultrarelativistic limit with $\omega_{k_x}>>
\omega_{k_z}$
\begin{equation}
\beta =\frac{1}{2}\omega_{k_x}
\end{equation}
and $-1/2\leq\lambda\leq 0$. 
The energy density is given by
\begin{equation}
<\rho>\simeq -\frac{\lambda\omega_{k_x}}{(1+\lambda^2)V}(\lambda+\frac
{1}{2}) .
\end{equation}
Thus in the limit $\omega_{k_x}/V\rightarrow\infty$ the energy density at the 
spacetime point $x^{\mu}=0$ goes to $-\infty$, for $-1/2<\lambda<0$.
   
Now consider, within the above limits, the energy density on the spacetime.
For general $x^{\mu}$
\begin{equation}
<\rho>=\frac{\lambda\omega_{k_x}}{(1+\lambda^2)V}
[\lambda+\frac{1}{2}\cos(\omega_{k_x}(t-x))].
\label{rho3}
\end{equation}
Thus $<\rho>$ is a cosine wave propagating at the speed of light
superimposed on a positive background. The energy density at a
fixed spatial point will be negative for a time $\Delta t$, which
satisfies
\begin{equation}
-V<\rho>_{min}\Delta t=\frac{|\lambda |(1+2\lambda)}{1+\lambda^2}\cos^{-1}
(2|\lambda |),
\end{equation}
where $<\rho>_{min}$ is the minimum value of $<\rho>$ for fixed $\omega_{k_x}$
and $\lambda$.
Since the wave propagates at the speed of light the extent of the negative
energy density will satisfy the same expression as above with $\Delta t$
replaced by $\Delta x$.
For large values of $|<\rho>|V$ (and $\lambda$ not too close to -1/2 or 0)
the energy density will undergo rapid
oscillations from positive to negative values. Note that the time average
of the energy density is positive.
\section*{Quantum Inequalities}
In this section I will show that the energy density satisfies the
quantum inequality (\ref{QI1}) for the state given in (\ref{state}).
To show this I will take the limit $m\rightarrow 0$ for the Dirac field.
Substituting (\ref{rho3}) into 
\begin{equation}
\hat{\rho}=\frac{t_0}{\pi}\int_{-\infty}^{\infty}\frac{<\rho>}
{t^2+t_0^2}dt
\end{equation}
and taking $\vec{x}=0$ gives
\begin{equation}
\hat{\rho}=\frac{\lambda\omega_{k_x}}{(1+\lambda^2)V}
[\lambda+\frac{1}{2}e^{-\omega_{k_x}t_0}].
\label{hat}
\end{equation}
Thus $\hat{\rho}$ will be negative if
\begin{equation}
\omega_{k_x}t_0<-\ln(2|\lambda |).
\end{equation}
Now consider the quantum inequality (\ref{QI1}) for A=1.
 
If periodic boundary conditions are imposed
\begin{equation}
\omega=\frac{2\pi}{L}\sqrt{n_x^2+n_y^2+n_z^2}
\end{equation}
where $n_x, n_y,$ and $n_z$ are integers. Next note that
\begin{equation}
\frac{1}{\sqrt{3}}(|n_x|+|n_y|+|n_z|)\leq\sqrt{n_x^2+n_y^2+n_z^2}\leq
|n_x|+|n_y|+|n_z|.
\label{ineq}
\end{equation}
To see this define $f(n_x,n_y,n_z)$ by
\begin{equation}
f(n_x,n_y,n_z)=\sqrt{n_x^2+n_y^2+n_z^2}-\alpha(|n_x|+|n_y|+|n_z|)
\label{f}
\end{equation}
where $\alpha$ is a positive constant. Now let
\begin{equation}
\begin{array}{ll}
|n_x|=r\cos\phi\sin\theta\\
|n_y|=r\sin\phi\sin\theta\\
|n_z|=r\cos\theta ,
\end{array}
\end{equation}
where $0\leq(\theta,\phi)\leq\pi/2$. Thus $f$ can be written as
\begin{equation}
f(r,\theta,\phi)=r(1-\alpha g(\theta,\phi))
\end{equation}
where
\begin{equation}
g(\theta,\phi)=\sqrt{1+2\sin^2(\phi+\pi/4)}\sin(\theta+\psi)
\end{equation}
and 
\begin{equation}
\cot(\psi)=\sqrt{2}\sin(\phi+\pi/4).
\end{equation}
Given that $0\leq(\theta,\phi)\leq\pi/2$, it is easy to see that $1\leq
g(\theta,\phi)\leq\sqrt{3}$. The function $g(\theta,\phi)$ has its
maximum value of $\sqrt{3}$ when $\phi=\pi/4$ and $\theta=\pi/2-\cot^{-1}(\sqrt{2})$. The
minimum value of $g(\theta,\phi)$ occurs when $\theta=0$. In this case
$g(0,\phi)=1$ for all $\phi$. The function $g(\theta,\phi)$ also equals 1
at $\theta=\pi/2$, $\phi=0,\pi/2$.
Now if $\alpha=1$ in (\ref{f}) then 
$f=r(1-g)\leq 0$. This gives the inequality $\sqrt{n_x^2+n_y^2+n_z^2}
\leq |n_x|+|n_y|+|n_z|$. On the other hand, if $\alpha=1/\sqrt{3}$
then $f=r(1-g/\sqrt{3})\geq 0$. This gives the inequality $3^{-1/2}
[|n_x|+|n_y|+|n_z|]\leq\sqrt{n_x^2+n_y^2+n_z^2}$. Thus (\ref{ineq})
is proved.
  
To show that (\ref{QI1}) is satisfied I will show that an even
more restrictive inequality is satisfied. From (\ref{ineq})
\begin{equation}
\omega_k^-\leq\omega_k\leq\omega_k^+
\end{equation}
where
\begin{equation}
\omega_k^-=\frac{2\pi}{\sqrt{3}L}[|n_x|+|n_y|+|n_z|]
\end{equation}
and
\begin{equation}
\omega_k^+=\frac{2\pi}{L}[|n_x|+|n_y|+|n_z|].
\end{equation}
Since
\begin{equation}
\omega_k^-e^{-2\omega_k^+t_0}\leq\omega_ke^{-2\omega_kt_0}
\end{equation}
th quantum inequality (\ref{QI1}) will be satisfied if
\begin{equation}
\hat{\rho}\geq-\frac{1}{2V}\sum_k\omega_k^-e^{-2\omega_k^+t_0}
\end{equation}
is satisfied. Substituting in the expressions for $\omega_k^+$ and
$\omega_k^-$ gives
\begin{equation}
\hat{\rho}\geq-\frac{2\sqrt{3}
\pi}{LV}\left[\sum_{n=0}^{\infty}ne^{-\alpha n}
\right]\left[2\sum_{k=0}^{\infty}e^{-\alpha k}-1\right]^2,
\end{equation}
where $\alpha=4\pi t_0/L$.
The sums can easily be performed, giving
\begin{equation}
\hat{\rho}\geq -\frac{2\sqrt{3}\pi}{LV}\frac{e^{-\alpha}(1+
e^{-\alpha})^2}
{(1-e^{-\alpha})^4}.
\end{equation}
Thus (\ref{QI1}) will be satisfied if 
\begin{equation}
\frac{\lambda n_x}{1+\lambda^2}(\lambda+\frac{1}{2}e^{-\alpha n_x/2})\geq
-\frac{\sqrt{3}e^{-\alpha}(1+e^{-\alpha})^2}
{(1-e^{-\alpha})^4}.
\label{45}
\end{equation}
is satisfied.
For $\lambda$ outside the interval $(-\frac{1}{2}e^{-\alpha n_x/2},0)$ the
above inequality will obviously be satisfied. Thus consider $-\frac{
1}{2}e^{-\alpha
n_x/2}<\lambda<0$. Equation (\ref{45}) will be satisfied if
\begin{equation}
|\lambda|n_x(\lambda+\frac{1}{2}e^{-\alpha n_x/2})\leq\frac{\sqrt{3}e^{-\alpha}
(1+e^{-\alpha})^2}{(1-e^{-\alpha})^4}
\end{equation}
is satisfied. Now let $\lambda=-\frac{\sigma}{2}e^{-\alpha n_x/2}$. The above
inequality becomes
\begin{equation}
\frac{1}{4}\sigma n_xe^{-\alpha n_x}(1-\sigma)\leq\frac{\sqrt{3}e^{-\alpha}
(1+e^{-\alpha})^2}{(1-e^{-\alpha})^4}.
\label{47}
\end{equation}
The left hand side is maximized for $\sigma=1/2$ and for $n_x=1/\alpha$.
But $n_x$ is a positive integer. Thus for $\alpha\geq 1$ take $n_x=1$. The
above inequality will then be satisfied (for $\alpha \geq 1$) if
\begin{equation}
\frac{1}{16}\leq\frac{\sqrt{3}(1+e^{-\alpha})^2}{(1-e^{-\alpha})^4}
\end{equation}
is satisfied. This is obviously satisfied for all $\alpha\geq 1$. For
$\alpha <1$ let $n_x=\frac{1}{\alpha}$ (i.e. generalize $n_x$ to a real
number). The inequality (\ref{47}) will be satisfied if
\begin{equation}
\frac{1}{16}e^{-1}\leq\frac{\sqrt{3}\alpha e^{-\alpha}(1+e^{-\alpha})^2}{(1-
e^{-\alpha})^4}
\end{equation}
is satisfied. In the interval $0<\alpha<1$ the right hand side is a
monotonically decreasing function of $\alpha$ with a minimum value of
$e^{-1}(1+e^{-1})^2(1-e^{-1})^{-4}$. Thus the above inequality is satisfied.
Therefore inequality (\ref{45}) will be satisfied, which implies that the
quantum inequality (\ref{QI1}) will be satisfied.
\section*{The Klein-Gordon Field}
In this section I will show that the energy density for a massive scalar field
is positive for all states that are arbitrary superpositions of single 
particle states.
  
The scalar field operator can be written in terms of creation and 
annihilation operators as
\begin{equation}
\phi(x)=\imath\sum_k\frac{1}{\sqrt{2V\omega_k}}(a_ke^{\imath k^{\mu}
x_{\mu}}-a_k^{\dagger}e^{-\imath k^{\mu}x_{\mu}}).
\end{equation}
The energy-momentum tensor for the scalar field is given by
\begin{equation}
T^{\mu\nu}=\partial^{\mu}\phi\partial^{\nu}\phi-\frac{1}{2}\eta^{\mu\nu}
(\partial_{\alpha}\phi\partial^{\alpha}\phi+m^2\phi^2).
\end{equation}
A short calculation gives
\begin{equation}
<\rho>=\frac{1}{2V}Re\sum_{k,k^{'}}\frac{(\omega_k\omega_{k^{'}}
+\vec{k}\cdot\vec{k}^{'}+m^2)}{\sqrt{\omega_k\omega_{k^{'}}}}
<a^{\dagger}_{k^{'}}a_k>e^{\imath(k^{\mu}-k^{'\mu})x_{\mu}}
\end{equation}
\begin{equation}
+\frac{(\omega_k
\omega_{k^{'}}+\vec{k}\cdot\vec{k^{'}}-m^2)}{\sqrt{\omega_k\omega_{
k^{'}}}}<a_{k^{'}}a_k>e^{\imath(k^{\mu}+k^{'\mu})x_{\mu}}
\end{equation}
The state vector can be written as
\begin{equation}
|\psi>=\frac{1}{N}\sum_k\alpha_k|k>
\label{state5}
\end{equation}
where the $\alpha_k$ are arbitrary complex numbers and $N$ is chosen so that
$|\psi>$ is normalized. The energy density can now be written as
\begin{equation}
<\rho>=\frac{1}{2VN^2}\sum_{k,k^{'}}\frac{(\omega_k\omega_{k^{'}}
+\vec{k}\cdot\vec{k}^{'}+m^2)}{\sqrt{\omega_k\omega_{k^{'}}}}
\alpha^*_{k^{'}}\alpha_k e^{\imath(k^{\mu}-k^{'\mu})x_{\mu}}.
\end{equation}
Now define
\begin{equation}
\beta=\sum_k\sqrt{\omega_k}\alpha_k e^{\imath k^{\mu}x_{\mu}},\;\;\;\;\;
\lambda=m\sum_k\frac{\alpha_k}{\omega_k} e^{\imath k^{\mu}x_{\mu}},
\end{equation}
and
\begin{equation}
\vec{\gamma}=\sum_k\frac{\vec{k}}{\sqrt{\omega_k}}\alpha_k e^{\imath k^{\mu}
x_{\mu}}.
\end{equation}
Thus for sate vectors of the form (\ref{state5})
\begin{equation}
<\rho>=\frac{1}{2VN^2}(|\beta|^2+|\lambda|^2+|\vec{\gamma}|^2),
\end{equation}
and the energy density is non-negative.
\section*{Conclusion}
In this paper I examined the negative energy densities that can be
produced in the Dirac field by state vectors of the form
\begin{equation}
|\psi>=\frac{1}{\sqrt{1+\lambda^2}}(|k_z,1>+\lambda|k_x,2>),
\label{conc}
\end{equation}
where $|k_z,1>$ and $|k_x,2>$ are single particle electron states
and $\lambda$ is real. I showed that if $k_x,k_z>>m$, the energy density
at a space-time point $x^{\mu}$ will be negative if $\lambda$ is
chosen so that
\begin{equation}
-\frac{\beta}{2}-\sqrt{\left(\frac{\beta}{2}\right)^2-\omega_{k_x}\omega_{k_z}}
<\omega_{k_x}\lambda <-\frac{\beta}{2}+\sqrt{\left(\frac{\beta}{2}\right)^2
-\omega_{k_x}\omega_{k_y}}
\end{equation}
is satisfied, where
\begin{equation}
\beta=\frac{1}{2}(\omega_{k_x}+\omega_{k_z})\cos[(k^{\mu}-k^{'\mu})x_{\mu}].
\end{equation}
Since I am taking $\lambda$ to be real it is necessary that $\beta\geq 4
\omega_{k_x}\omega_{k_z}$. This will be satisfied if $\omega_{k_x}\geq
(7+\sqrt{48})\omega_{k_z}$ or if $\omega_{k_z}\geq (7+\sqrt{48})\omega_{k_x}$.
  
If, in addition to $k_x,k_z>>m$, one takes $\omega_{k_x}>>\omega_{k_z}$
then
\begin{equation}
-\frac{1}{2}\leq\lambda\leq 0\;,\;\;\;\;\;\;\;\; \beta=\frac{1}{2}
\omega_{k_x}
\end{equation}
and
\begin{equation}
<\rho>=\frac{\lambda\omega_{k_x}}{(1+\lambda^2)V}(\lambda+\frac{1}{2})
\end{equation}
at the point $x^{\mu}=0$. Thus $<\rho>\rightarrow -\infty$ as 
$\omega_{k_x}/V\rightarrow\infty$, for $-1/2<\lambda<0$ and $<\rho>$
is not bounded from below.
   
An observer will see $<\rho>$ as a cosine wave propagating at the
speed of light superimposed on a positive background. The time average
of $<\rho>$ is positive and the energy density will be negative for a
time interval $\Delta t$, which satisfies
\begin{equation}
-V<\rho>_{min}\Delta t=\frac{|\lambda|(1+2\lambda)}{1+\lambda^2}\cos^{-1}
(2|\lambda|).
\end{equation}
I also showed that the quantum inequality
\begin{equation}
\hat{\rho}\equiv\frac{t_0}{\pi}\int_{-\infty}^{\infty}\frac{<\rho>dt}
{t^2+t_0^2}\geq-\frac{1}{2V}\sum_k\omega_k e^{-2\omega_k t_0},
\end{equation}
which is satisfied by a massless scalar field, is satisfied by the
Dirac field for state vectors of the form (\ref{conc}) in the limit
$m\rightarrow 0$. Finally, I showed that, in contrast to the Dirac
field, it is not possible to produce negative energy densities in a scalar
field using state vectors that are arbitrary superpositions of single
particle states.

\end{document}